\documentclass[aps,tightenlines,nofootinbib,preprint,groupedaddress]{revtex4}
\usepackage{graphicx}
\newcommand{\be}{\begin{equation}}
\newcommand{\ee}{\end{equation}}
\newcommand{\bear}{\begin{eqnarray}}
\newcommand{\eear}{\end{eqnarray}}
\begin{document}

\title{Direct resummation of the leading renormalons in inclusive \\
semileptonic B decay}

\author{Taekoon Lee}
\email{tlee@muon.kaist.ac.kr}

\affiliation{Department of Physics, Korea Advanced Institute of
Science and Technology, Daejon 305-701, Korea}


\begin{abstract}
We present a Borel resummation method for the QCD corrections in
inclusive, charmless, semileptonic B meson decay.
The renormalon contributions are resummed to all orders 
by employing a bilocal expansion of the Borel 
transform that accurately accounts for the first infrared 
renormalons in the Borel plane. The renormalons in the pole mass and
the QCD expansion are resummed separately, and a 
precise relation is obtained between a properly defined pole mass and 
the ${\overline {\rm MS}}$ mass.
The inclusive decay rate is calculated to three loop order
using an estimate of the yet unknown NNLO coefficient.
\end{abstract}



\maketitle


 The Cabibbo-Kobayashi-Maskawa matrix element $|V_{\rm ub}|$,
 which is an important parameter in the study of CP violation in
 weak interactions, can be determined through the inclusive,
 charmless, semileptonic B decay. Presently, the largest theoretical 
 uncertainty in the decay rate comes from the $b$ quark mass and the
 QCD correction. It is well known that these two are not independent but 
 intimately related. 
 In the large $b$ quark
 mass limit, the B meson decay rate is given by that of a free
 $b$ quark with a correction quadratically suppressed in the $b$ mass
 \cite{decay1,decay2}.
 The decay rate in the on-shell scheme is given in the form
 \be
 \Gamma(B\to X_{ue\bar\nu_e}) =\frac{G_{\rm F}^2|V_{\rm ub}|^2}{192\pi^3}
 m_b^5 f(\alpha_s)[1+ O(1/m_b^2)]
\label{eq1}
\ee
 with $m_b$ and  $f(\alpha_s)$ denoting
 the $b$ quark pole mass and the QCD correction to the free $b$ quark
decay, respectively.
In perturbation theory $f(\alpha_s)$ can be expanded as
\be
f= 1+\sum_{n=0}^\infty f_n(\xi) \alpha_s(\mu)^{n+1}\,,
\label{e2}
\ee
where $\alpha_s(\mu)$ is the strong coupling constant  and
$\xi\!=\!\mu/m_{\overline {\rm MS}}$, with $m_{\overline {\rm MS}}
[\equiv m_{\overline {\rm MS}}(m_{\overline {\rm MS}})]$ denoting
the  $\overline {\rm MS}$  mass.

Because of the infrared (IR) renormalon the perturbative coefficients $f_n$
are expected to grow rapidly, and this would give
rise to an  $O(\Lambda_{\rm QCD}/m_b)$  intrinsic uncertainty
to the asymptotic series. On the other hand, this renormalon uncertainty in
$f(\alpha_s)$ is known to arise from the use of the pole mass in the
decay rate. When the pole mass is expanded in the strong coupling
it also has a renormalon caused
uncertainty of $O(\Lambda_{\rm QCD})$, 
and this uncertainty cancels that of
$f(\alpha_s)$ \cite{bigi,bz}. Once the pole mass is replaced by 
a renormalon free mass like the $\overline{\rm MS}$ mass or one of the
threshold masses \cite{kin,ps,1smass,1smass1}
the resulting $f(\alpha_s)$ is  in principle 
expected to give a well-behaved series.
In practice, however, the use of the renormalon free masses are not completely
satisfactory. For instance, with the ${\overline {\rm MS}}$ mass the
renormalon cancellation is not so obvious at low orders and the convergence
for $f(\alpha_s)$ is quite slow, possibly due to the large power of $m_b$
in the decay rate, and with the $1S$ mass \cite{1smass,1smass1}
one must assume that the $1S$ state of $\Upsilon$ is a perturbative system.

In this paper we describe a more direct approach to the renormalon problem,
in which the leading renormalons in the pole mass and $f(\alpha_s)$ are
Borel resummed separately. An essential ingredient of this program
is the accurate calculation of the renormalon residue for the pole mass. 
The residue, which determines the normalization constant of the
large order behavior, can be computed in perturbation theory
\cite{lee-residue1,lee-residue2}, and 
it turns out that the residue for the pole mass  can be determined to good
accuracy within a few percent error \cite{lee-br,pineda1,pineda2}.
The renormalon cancellation then allows
the residue for $f(\alpha_s)$  to be determined to the same accuracy from 
the pole mass residue. The Borel transforms of the pole mass and $f$ can be
systematically expanded around the origin and the first renormalon location in
the Borel plane.
With the computed residues these two expansions
can be interpolated to a {\it bilocal expansion} to obtain an
accurate description of the Borel transforms in the region in-between the 
two expansion points. Since at the typical strong coupling in B decay 
the bulk of the Borel integral comes from  this region of integration, 
the bilocal expansion can much improve the accuracy of the Bore resummation.
By taking into account the renormalon singularity
correctly this program essentially resums the renormalon caused 
large order behavior to all orders.

With the renormalon singularity on the integration contour one has to worry
about  the ambiguities of  $O(\Lambda_{\overline{\rm MS}})$ for
the pole mass and $O(\Lambda_{\overline{\rm MS}}/m_b)$ for $f$  
arising from the 
ambiguities in choosing the integration contour that may either be on the upper
or the lower half plane beginning at the renormalon singularities.
It was shown in Refs. \cite{luke,neubert} that these ambiguities cancel
as long as one takes the same contour for the pole mass 
and $f(\alpha_s)$. Furthermore the authors argued that in general
the pole mass, when consistently used, 
does not cause any ambiguity to physical observables
in the heavy quark effective theory.
Until recently, however, there was no procedural program implementing
this observation. The main reason for this, perhaps, was that one did not 
know how to describe
the Borel transforms beyond the immediate
neighborhood of the origin in the Borel
plane. Within the large $\beta_0$ approximation \cite{largebeta}
one can obtain the Borel
transforms in the whole Borel plane, but the accuracy of the approximation is
limited. What is necessary for practical applications is an accurate
description of the Borel transforms in the region in-between the origin 
and the first renormalon singularity, since the contribution
from the region far from the origin is exponentially suppressed.
The bilocal expansion fills this gap.

The program described above was successfully applied  to the 
heavy quark static potential \cite{lee-br} which is known to suffer from 
a severe convergence 
problem with the ordinary perturbative expansion. 
The Borel resummed potential
obtained from employing a bilocal expansion turns out to have an excellent
convergence and is in remarkable agreement with the lattice calculations.

Throughout the paper, unless stated otherwise, the perturbative expansions
are assumed to be in the ${\overline {\rm MS}}$ scheme with four
active quark flavors ($N_f\!=\!4$), and only the leading IR renormalon 
closest to the origin in the Borel plane is considered.

The pole mass in terms of the ${\overline{\rm MS}}$
mass $m_{\overline{\rm MS}}$
has the perturbative expansion
\be
m_b=m_{\overline{\rm MS}} \left[ 1+\sum_{n=0}^{\infty} p_n(\xi) 
\alpha_s(\mu)^{n+1} \right]\,.
\label{e3}
\ee
The first three coefficients $p_n$ 
are known at $\mu=m_{\overline{\rm MS}}$ \cite{massexpansion1,
massexpansion2,massexpansion3} and their $\mu$ dependence 
can be obtained by the renormalization group  invariance of 
the pole mass.
The Borel resummation of this expansion can be written formally as
\bear
m_{b}[\alpha_s(\mu)] 
=m_{\overline {\rm MS}}\left[ 1 +\frac{1}{\beta_0}
\int_{0\pm i\epsilon}^{\infty\pm i\epsilon}
e^{-b/\beta_0\alpha_s(\mu) } \tilde m(b,\xi) \,db\right]\,,
\label{e5}
\eear
where the  Borel transform $\tilde m(b,\xi)$ has
the perturbative expansion
\be
\tilde m(b,\xi)=\sum_{n=0}^\infty \frac{p_n(\xi)}{n!}
\left(\frac{b}{\beta_0}\right)^n\,,
\label{m-boreltransform}
\ee
with $\beta_0$ denoting the one loop coefficient of the QCD $\beta$ function
[$\beta\!=\!d\alpha_s/d\ln(\mu^2)\!=\!-\alpha_s^2(
\beta_0+\beta_1\alpha_s+\beta_2\alpha_s^2+{\ldots})]$.
The integration contour, which can be derived from the general argument
of Borel resummation of a series with a same sign large order behavior
\cite{lee-np},
 can be either on
the upper or the lower half plane along the positive real axis.
The pole mass is known to have an ambiguity from renormalon singularity 
proportional to $\Lambda_{\overline {\rm MS}}$ 
\cite{beneke1} that has the weak coupling 
expansion
\bear
 \Lambda_{\overline{\rm MS}} 
\propto \mu\alpha_s(\mu)^{-\nu}
e^{-\frac{1}{2\beta_0\alpha_s(\mu)}} \left[1
+ r_1 \alpha_s(\mu) +r_2\alpha_s(\mu)^2
+{\ldots} \right] \,,
\label{e7}
\eear
where
\bear
\nu &=& \frac{\beta_1}{2\beta_0^2}\,, \hspace{0.25in} r_1=\frac{
\beta_1^2-\beta_0\beta_2}{2\beta_0^3}\,, \nonumber\\
r_2&=& \frac{
 \beta_1^4 +4\beta_0^3\beta_1\beta_2
-2 \beta_0\beta_1^2\beta_2
+\beta_0^2(\beta_2^2-2\beta_1^3)-2\beta_3\beta_0^4}{
8\beta_0^6} \,.
\label{e8}
\eear
For the Borel integral (\ref{e5}) to generate this ambiguity
one can easily see that the Borel
transform $\tilde m(b,\xi)$ must have the singularity 
\bear
\tilde m(b,\xi) = \frac{C_m \xi}{(1-2b)^{1+\nu}}\left[
1+c_1
(1-2b)+c_2
(1-2b)^2+{\ldots}\right] + \text{ analytic part}\,,
\label{e9}
\eear
where
\be 
c_1=\frac{r_1}{2\nu\beta_0}, \hspace{.5in} c_2=
\frac{r_2}{4\nu(\nu-1)\beta_0^2} \,.
\ee
The `analytic part' denotes any terms analytic around $b=1/2$ with radius of
convergence bounded by the next renormalon at $b=3/2$ and
$C_m$ is a real constant.

With this singularity  in $\tilde m(b,\xi)$ the resummed pole mass can be
written formally as
\be
m_b= m_{\rm BR}[\alpha_s(\mu)] \pm i \Gamma_{m}[\alpha_s(\mu)]\,,
\ee
where the `Borel resummed (BR)' mass $m_{\rm BR}$ denotes
the real part of the Borel integral in Eq. (\ref{e5})
and $\Gamma_{m}$ denotes the imaginary part obtained with the integration 
contour on
the upper half plane.
Note that by definition the BR mass has the same perturbative expansion
as the pole mass and is not a short distance mass.

The residue $C_m$, which is not known exactly, can be calculated
systematically  in 
perturbation theory \cite{lee-residue1,lee-residue2}. The details of
the calculation adopted here can be found in \cite{lee-br}.
To compute the residue we first expand using the known $p_n$ up to NNLO 
and the unknown $p_3$  the function
$
R(b,\xi)\equiv(1-2b)^{(1+\nu)}\tilde m(b,\xi)
$
about the origin in the $w$ plane defined by the following conformal
mapping
\be
w=\frac{\sqrt{1+b}-\sqrt{1-2b/3}}{\sqrt{1+b}+\sqrt{1-2b/3}}\,,
\ee
which gives to $O(w^3)$
\be
R[b(w),1]= 0.42441+0.61198 w+0.25351 w^2+[-108.99217+7.90052\, p_3(1)] w^3\,.
\label{rbw}
\ee
This allows us to estimate $p_3$ \cite{lee-coef}. 
From the pattern of the known lower order terms 
we may safely assume
$
|108.99217-7.90052\, p_3(1)|<1\,,
$
which leads to
\be
p_3(1)=13.796\pm 0.127\,.
\label{p3estimate}
\ee
Now evaluating Eq. (\ref{rbw}) at the renormalon location $w\!=\!1/5$ we
obtain the residue
\be
C_m=R[b(w\!=\!1/5),1]\approx
0.42441+0.12240+0.01014 \pm 0.00800=0.55695\pm 0.00800\,,
\label{cm}
\ee
where the error comes from the uncertainty in the estimate (\ref{p3estimate}).

The BR mass $m_{\rm BR}$ can be obtained accurately
by an interpolation of the two expansions (\ref{m-boreltransform})
and (\ref{e9}) of the Borel transform
$\tilde m(b,\xi)$ (bilocal expansion \cite{lee-br}).
Since at the typical values of the strong coupling considered here
the bulk of the contribution 
to the Borel integral comes from the
region between the origin and the first IR renormalon, it is important
to have the Borel transform in this region as accurately as possible
from the known first perturbative terms. As demonstrated  
in the heavy quark potential
the accuracy of the Borel transform can be improved greatly through the
bilocal expansion  \cite{lee-br}.

The bilocal expansion of the Borel transform for the pole mass is given by
\bear
\tilde m(b,\xi) &=&\lim_{N,M \to \infty} \tilde m_{\rm N,M} 
(b,\xi) \nonumber\\
&=& \lim_{N,M \to \infty}\left\{
\sum_{n=0}^N\frac{q_n(\xi)}{n!} \left(\frac{b}{\beta_0}\right)^n
+\frac{C_m\xi}{(1-2b)^{1+\nu}}
\left[ 1 +\sum_{i=1}^M c_i (1-2b)^i\right]\right\}
\label{bilocal} \,,
\eear
where $q_n$ are to be determined by demanding Eq. (\ref{bilocal})
reproduce Eq. (\ref{m-boreltransform}) when expanded about  $b=0$.
With the known first two coefficients $c_{1,2}$, and taking 
$M=2$, we  have the first three $q_n$ as
\bear
q_0(\xi)&=& p_0 -C_m\xi(1+c_1+c_2) \,, \nonumber\\
q_1(\xi)&=&p_1(\xi) -2C_m\xi\beta_0[1-c_2+\nu(1+c_1+c_2)] \,,\nonumber\\
q_2(\xi)&=& p_2(\xi)-4C_m\xi\beta_0^2[2+\nu(3+c_1-c_2)+\nu^2(1+c_1+c_2)]\,.
\eear
Since the two expansions about the origin and about the renormalon
singularity are expected to be convergent on the disks, $|b| < 1/2$
and $|b-1/2|< 1$, respectively, the interpolation (\ref{bilocal})
is expected to give a good description of the Borel transform in the
region that includes the two expansion points.

Taking the canonical value for the strong coupling 
$\alpha_s(m_{\overline{\rm MS}})= 0.22$ and $\xi\!=\!1$, 
and substituting
the known $\tilde m_{\rm N,2} (b,1), (N=0,1,2)$, into the Borel
integral (\ref{e5}), and performing the integration in the $w$-plane
for the convenience of numerical integration, we obtain 
\be
m_{\rm BR}= m_{\overline{\rm MS}} (1+0.15769+0.00409-0.00028\mp 0.00014)\,,
\label{mbr}
\ee
where the error was obtained by varying $C_m$ within 
the uncertainty in the computed  residue 
in Eq. (\ref{cm}).
Notice the remarkable convergence of the BR mass compared to the power 
series (\ref{e3}) which at $\xi=1$ gives 
$m_b=m_{\overline{\rm MS}}(1+0.09337+0.04550+0.03235)\,. $
This shows that the bilocal expansion improves the convergence by
two orders of magnitude.

We now turn to the resummation of $f(\alpha_s)$.
For this we first note that because of the large power ($m_b^5$) in
the pole mass
in the decay rate the expansion (\ref{e2}) may not be an optimal
expansion for $f(\alpha_s)$.
We can reorganize the expansion  (\ref{e2}) by writing
\be
f[\alpha_s(\mu)]=
F\biglb(1+H[\alpha_s(\mu)]\bigrb)
\label{e14}
\ee
where $F(x)$ is an arbitrary function analytic at $x=1$.
Since $H$ itself can be regarded as a coupling constant, with its relation to
the $\overline {\rm MS}$ coupling constant defined implicitly by 
Eq. (\ref{e14}), we see that
a particular choice of $F$ is nothing but a renormalization scheme
choice.

Given an $F$ the perturbative expansion for $H$
\be
H= \sum_{n=0}^\infty h_n (\xi) \alpha_s(\mu)^{n+1}
\label{e15}
\ee
can be 
easily obtained using the definition (\ref{e14}) and the series (\ref{e2}) for
$f(\alpha_s)$. Instead of resumming the series (\ref{e2}) directly we shall
resum  the expansion (\ref{e15}) for $H$.

The Borel resummation for $H$ can be done in a similar fashion as in
the pole mass resummation, assuming that the first IR renormalon for $H$
is known.
Then with the resummed pole mass and  $H$
we have
\be
m_b^5 f= (m_{\rm BR}  \pm i \Gamma_m)^5
F\biglb(1+H_{\rm BR} \pm i \Gamma_H)\,,
\label{e16}
\ee
where
$H_{\rm BR}$
denotes the real part of the Borel integral 
\be
H[\alpha_s(\mu)]=\frac{1}{\beta_0}
\int_{0\pm i\epsilon}^{\infty \pm i\epsilon}
e^{-b/\beta_0\alpha_s(\mu) } \tilde H(b,\xi) \,db 
\label{e17}
\ee
 and $\Gamma_H$ denotes the imaginary part obtained with the contour on the
 upper half plane.
The Borel transform $\tilde H(b,\xi)$ has the perturbative expansion
\be
\tilde H(b,\xi)=\sum_{n=0}^\infty 
\frac{h_n(\xi)}{n!}\left(\frac{b}{\beta_0}\right)^n\,.
\label{h-boreltransform}
\ee

The structure of the IR renormalon for $H$  can be easily obtained
by the cancellation of the renormalons in $m_b^5$ and  $f(\alpha_s)$.
Expanding Eq. (\ref{e16}) in powers of $\Gamma_{m,H}$ we obtain
\be
m_b^5 f=m_{\rm BR}^5 F(1+H_{\rm BR})
\pm i\, m_{\rm BR}^4 \left[5 F(1+H_{\rm BR}) \Gamma_m
+F'(1+H_{\rm BR})m_{\rm BR}
\Gamma_{H}\right] +O(\Gamma_{m, H}^2)
\label{e19}
\ee
The cancellation of the renormalons 
means vanishing of the ambiguous imaginary part, which gives
\be
\Gamma_H(\alpha_s)=-\frac{5 F[1+H_{\rm BR}(\alpha_s)]}
{F'[1+H_{\rm BR}(\alpha_s)]m_{\rm BR}(\alpha_s)}  
\Gamma_m(\alpha_s) \,.
\label{e20}
\ee
The renormalon singularity for $\tilde H(b,\xi)$
can now be easily obtained by expanding the right-hand-side
of this equation in small $\alpha_s$.

At this point we choose a specific form for $F(x)$.
Considering the large power of the  pole mass in the decay rate
a plausible choice for $F$ is  
\be
F(x)=\left[1+\frac{1}{q}(x-1)\right]^q
\label{fx} \,,
\ee
with q a number not necessarily an integer.  $F(1+H)$  is then essentially 
a power function with a rescaling of
the  normalization for $H$ that is for the purpose of convenience only and
does not have any physical effect.

Using the perturbative expansions in Eqs. (\ref{e3}) and (\ref{e15}) 
of $m_{\rm BR}$ 
and $H_{\rm BR}$ (which have
the identical expansions with $m_b$ and $H$, respectively) 
and $\Gamma_m$
obtained from the renormalon singularity (\ref{e9}), 
we have 
\bear
\Gamma_H(\alpha_s)=
\frac{5 C_m \xi \Gamma(-\nu)\sin(\nu\pi)}{
(2\beta_0)^{1+\nu}}\alpha_s^{-\nu}
e^{-\frac{1}{2\beta_0\alpha_s}} \left[1
+ \tilde{r}_1 \alpha_s +\tilde{r}_2(\xi)\alpha_s^2
+{\ldots} \right] \,,
\eear
where 
\bear
\tilde {r}_1&=&r_1 -p_0+h_0/q \nonumber\\
\tilde {r}_2(\xi)&=&r_2-p_1(\xi)+p_0(p_0-r_1)+[h_1(\xi)+h_0 r_1-p_0 h_0]/q\,,
\label{rtilde}
\eear
with $r_{1,2}$ given in Eq. (\ref{e8}).
For the Borel integral (\ref{e17}) to have this imaginary part the Borel
transform $\tilde H(b,\xi)$ must have the singularity
\bear
\tilde H(b,\xi) = \frac{C_H(\xi)}{(1-2b)^{1+\nu}}\left[
1+\tilde{c}_1
(1-2b)+\tilde{c}_2(\xi)
(1-2b)^2+{\ldots}\right] + \text{ analytic part}\,,
\label{e23}
\eear
with
\be 
C_H(\xi)=-5C_m\xi, \hspace{.5in}
\tilde{c}_1=\frac{\tilde{r}_1}{2\nu\beta_0}, \hspace{.5in}
\tilde{c}_2(\xi)=\frac{\tilde{r}_2(\xi)}{4\nu(\nu-1)\beta_0^2} \,.
\label{e24}
\ee

Now the decay rate in the BR scheme is given by
 \be
 \Gamma(b\to X_{ue\bar\nu_e}) =\frac{G_{\rm F}^2|V_{\rm ub}|^2}{192\pi^3}
 m_{\rm BR}^5 f_{\rm BR}(\alpha_s)[1+ O(1/m_{\rm BR}^2)]\,,
\label{rateinbr}
\ee
where $f_{\rm BR}= F(1+H_{\rm BR})$. Note that this is nothing
but the decay rate in the on-shell scheme with Borel resummation,
but now there is no
theoretical ambiguity inherent in the formula in Eq. (\ref{eq1}).
Unlike $f_{\rm BR}$, for instance,  $f(\alpha_s)$ in Eq. (\ref{eq1}) 
is not well defined due to the divergence of the expansion (\ref{e2}).

\begin{figure}
 \includegraphics[angle=90 , width=9cm
 ]{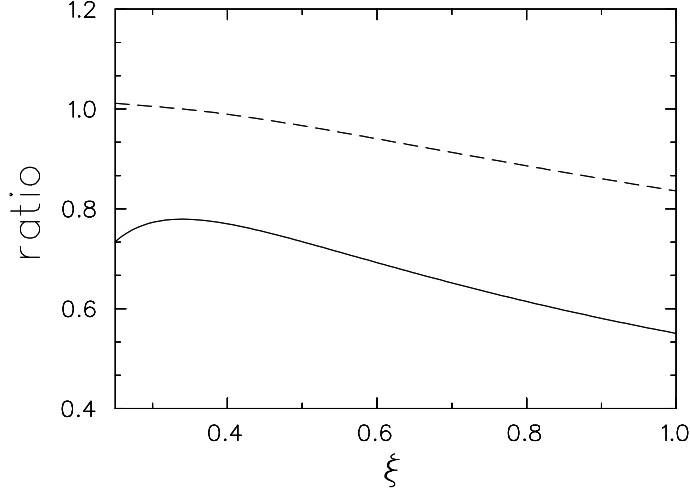}
\caption{\label{fig1} 
The ratio of the computed renormalon residue to the
expected one from the renormalon cancellation 
at NLO (solid) and at estimated NNLO (dashed) 
vs renormalization scale
$\xi\!=\!\mu/m_{\overline{\rm MS}}$.
}
\end{figure}

We now take the $q\to\infty$ limit in Eqs. (\ref{fx}) and (\ref{rtilde}),
which is equivalent to putting $f=\exp(H)$.
What is nice with this limit is
that the renormalon cancellation now occurs within the sum of
the series for $H$ and that of $5 \ln (m_b)$. We think this is
an attractive feature, considering the good cancellation
of renormalons within the sum of the pole mass and the static potential
in heavy quarkonium system \cite{pineda1,pineda2}.
Also, we shall see that the $q$ dependence of
$f(\alpha_s)$ becomes minimal in this limit.

Before starting computation of $H_{\rm BR}$ 
we first check how well the renormalon residue $C_H(\xi)$
computed using the known NLO calculation for $H$ compares to
the  one in Eq. (\ref{e24}) expected from the renormalon cancellation. 
This exercise will allow us to estimate 
the unknown NNLO coefficient for $f$.
The calculation for the residue $C_H(\xi)$ is completely
similar to that of the pole mass residue $C_m$.
The ratio of the calculated NLO residue to the expected residue 
is plotted in Fig. \ref{fig1}.
The best matching between the calculated and the expected
occurs at $\xi\approx0.34$ with about $80\%$ agreement.
This scale is 
smaller than $b$ quark mass, and this is in line with
the expectation that the optimal expansion for $f(\alpha_s)$
would occur at a scale somewhat below the $b$ mass.
This also shows that the NNLO contribution to the residue
should be significant. Now the unknown NNLO coefficient $f_2$
can be estimated  in the following way. We demand
the NNLO residue at  $\xi\!=\!0.34$ 
fall within a conservative 
 $20\%$  of the expected value ($|{\rm ratio} -1|\leq 0.2)$, which
 gives
 \be
 0.8\leq-1.552-0.277\, f_2(1)\leq1.2\,.
 \ee
 This provides us an estimate of the NNLO coefficient:
\be
f_2(1)=-9.22\pm0.72\,.
\label{e27}
\ee
This corresponds to  $f =[1-2.41 \bar a_s-21.30
\bar a_s^2 -(286\pm22)\bar a_s^3]$ 
and $m_b^5f=m_{\overline{\rm MS}}^5
[1+4.25\bar a_s+26.78\bar a_s^2+(160\pm
22)\bar a_s^3]$ where $\bar a_s=\alpha_s(m_{\overline{\rm MS}})/\pi$.
The latter shows that our estimate of the NNLO coefficient
sits in the middle of the estimate $5^3=125$
based on a naive power growing behavior \cite{bsuv}
and the Pad\'e estimate 188 \cite{steele}. 

\begin{figure}
\includegraphics [angle=90, width=9cm] {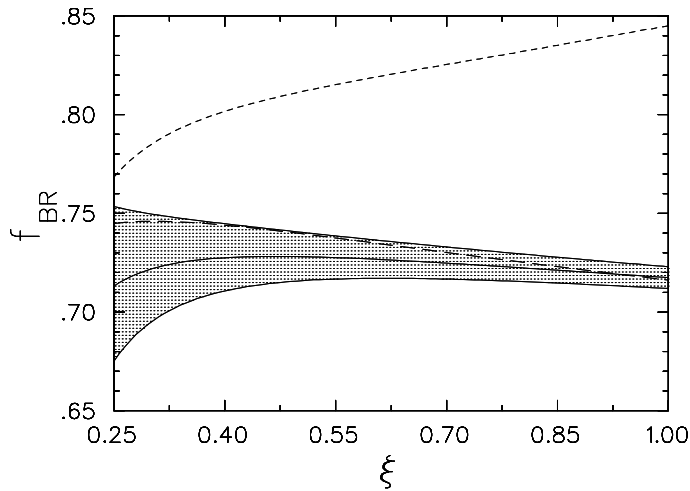}
\caption{\label{fig2} 
$f_{\rm BR}$ at LO (dotted), NLO (dashed), and estimated NNLO (solid) at
$\alpha_s(m_{\overline{\rm MS}})=0.22$ vs
renormalization scale $\xi\!=\!\mu/m_{\overline{\rm MS}}$.
The shaded band shows the uncertainty of $f_{\rm BR}$ at  NNLO.
}
\end{figure}

The computation of $H_{\rm BR}$ is completely parallel to
that of $m_{\rm BR}$. We first combine the two expansions 
in Eqs. (\ref{h-boreltransform}) and (\ref{e23})
for $H$ into the bilocal expansion $\tilde H_{N,M}(b,\xi)$ defined 
similarly to that of the pole mass in Eq. (\ref{bilocal}). The first two
$\tilde H_{N,2}(b,\xi)$ ($N=0,1$) can be obtained 
from the known NLO expansions \cite{ritgen}
of $f(\alpha_s)$, and the NNLO $\tilde H_{2,2}(b,\xi)$ from the
estimated coefficient in Eq. (\ref{e27}).
Substituting these into the Borel integral (\ref{e17}) and
performing the  integration 
the Borel resummed $H_{\rm BR}$ up to NNLO is obtained. The result for
$f_{\rm BR}=\exp{(H_{\rm BR})}$ is in Fig. \ref{fig2}.
The plots show that the scale dependence is significant at LO
but becomes milder at NLO and small at NNLO for $\xi\geq 0.5$.
Note that the NLO result is completely within the NNLO band
arising from the error in the $f_2$ estimate (\ref{e27}).
From the perspective of scale dependence the 
estimate in Eq. (\ref{e27}) with the
lower bound in the error estimate is preferred because it
defines the lower bound of the NNLO band which has a less scale 
dependence than the upper one.
In the presence of this unphysical scale dependence we can take the
principle of minimal sensitivity (PMS) \cite{pms}
values as the predictions of
the NNLO resummations. This gives 
\be
f_{\rm BR}=e^{H_{\rm BR}}= 0.728 \pm 0.011\,,
\label{e28}
\ee
where the error was taken from the difference between the PMS values
of the central line and the lower bound of the NNLO band.
Since an error estimate is always subjective, a more conservative
estimate can be obtained, for example, by taking the variation of the NNLO
estimates at $\xi\!=\!0.4$ as the uncertainty, 
which gives $f_{\rm BR}\approx 0.728 \pm 0.017$.

\begin{figure}
\includegraphics[angle=90 , width=9cm
 ]{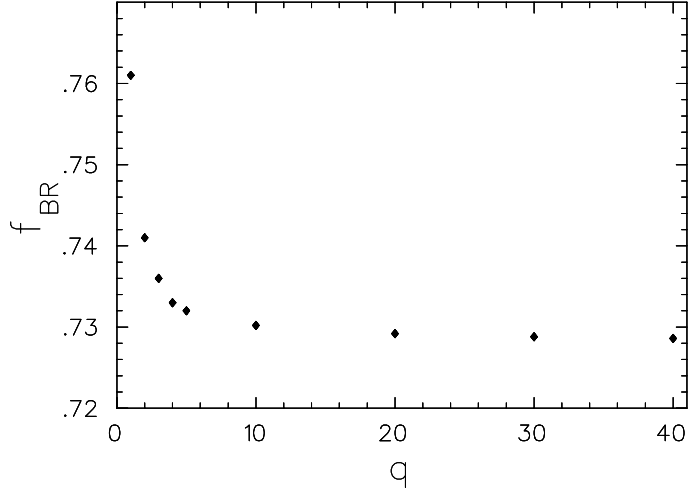}
\caption{\label{fig3} 
Renormalization scheme dependence of $f_{\rm BR}$ at NNLO (estimated).
}
\end{figure}

We now come back to the case of a finite value of $q$.
In principle $f_{\rm BR}(\alpha_s)$ should not depend on $q$, but
with a finite order perturbation there is always 
scheme dependence, and a scheme with a less 
scheme dependence is obviously preferable. 
To see the $q$ dependence we repeated the
NNLO computation  of $f_{\rm BR}$ at various values of $q$, and
the result is in Fig. \ref{fig3}.
The scheme dependence at small $q$ is quite significant,
for example, the difference between 
$q=1$ and $q=5$ is larger than the NNLO contribution at $\xi\!\geq\!0.25$. 
For large $q$ the scheme dependence is very mild and becomes
minimal at $q=\infty$.
This shows that
$q=1$, the usual choice in the literature,
is not a good scheme, but rather $q=\infty$ is  a good scheme
to choose.

Combining $m_{\rm BR}$ and $f_{\rm BR}$ in Eqs. (\ref{mbr}) and  (\ref{e28}),
respectively, we finally obtain the B decay rate  given 
in Eq. (\ref{rateinbr})
with
\be
m_{\rm BR}^5 f_{\rm BR}=m_{\overline {\rm MS}}^5
(1.539\pm 0.023)\,.
\ee

To conclude, we summarize our results as follows.
\begin{itemize}
{\item 
The leading renormalon in the pole mass can be resummed accurately
by employing a bilocal expansion.
The BR mass $m_{\rm BR}$, which is the real part of the integral (\ref{e5}),
is unambiguous and converges rapidly, although it has the same divergent
perturbative expansion as the pole mass and is not a short distance mass. 
The renormalon ambiguities in the pole mass
cancel when a consistent integration contour
is taken for all Borel integrals. Defined as the pole mass, 
the BR mass gives a precise meaning to
the relation between the pole mass and a high energy mass like the
$\overline{\rm {MS}}$ mass.}
{\item 
A leading renormalon in the effective field theories is no longer
a huddle but a blessing for the perturbation theory. By providing
a manageable constraint on the functional form of the Borel transform
it allows one to speed up the convergence of perturbation theory.}
{\item 
The renormalon residue of the pole mass can be determined accurately.
Using the renormalon cancellation one can determine the residue of
the other perturbative expansion that cancels the renormalon ambiguity
in the pole mass to the same accuracy. In the case of the B decay,
we used this information on the residue to estimate the NNLO 
coefficients $f_2$ in Eq. (\ref{e27}).}
{\item
The renormalization scheme dependence of $f(\alpha_s)$, parametrized
by Eq. (\ref{fx}),
is significant and should be taken into account.}
{\item
The B decay rate was calculated to NNLO (using an estimate of $f_2$)
in the BR scheme.}
\end{itemize}

\begin{acknowledgements}
This work was supported in part by BK21 Core Project.

\end{acknowledgements}


\end{document}